\documentclass[journal]{IEEEtran}
\usepackage{amsmath, amssymb, mathtools, bbm, graphicx, subfig, paralist}

\newtheorem{theorem}{Theorem}%[section]
\newtheorem{corollary}[theorem]{Corollary}
\newtheorem{remark}[theorem]{Remark}

\newcommand{ \out }[1]{ \operatorname{Output}({\bf #1}) }
\newcommand{ \mymod }{ \;\operatorname{mod}\, }
\newcommand{ \mybf }[1]{ \pmb{#1} }
\newcommand{ \myqed }{ \hfill $\blacktriangle$ }

\newcommand{ \defeq }{ \coloneqq }

\begin{document}

\title{On Error Detection in Asymmetric Channels}

\author{Mladen~Kova\v{c}evi\'c,~\IEEEmembership{Member,~IEEE}%
\thanks{Date: September 9, 2017.}
\thanks{%Manuscript received April 2, 2017; revised May 16, 2017.
        %The Associate Editor coordinating the review of this letter and approving it for publication was Hamid Saeedi.
        %
				This work was supported by the Singapore Ministry of Education (MoE) Tier 2 grant
        ``Network Communication with Synchronization Errors: Fundamental Limits and Codes"
        (Grant number R-263-000-B61-112).
				
        The author is with the Department of Electrical \& Computer Engineering,
        National University of Singapore %, Singapore 117583
				(email: mladen.kovacevic@nus.edu.sg).}
%\thanks{Digital Object Identifier}%
       }%

% \subjclass[2010]{Primary: 94B05, 94B25, 68P30;
%                  Secondary: 05D99, 11T71, 68R05, 52C17, 94B65.}

\maketitle

\begin{abstract}
  We study the error detection problem in $ \mybf{q} $-ary asymmetric channels
wherein every input symbol $ \mybf{x_i} $ is mapped to an output symbol $ \mybf{y_i} $
satisfying $ \mybf{y_i \geq x_i} $.
A general setting is assumed where the noise vectors are (potentially) restricted in:
\begin{inparaenum}[1)]
\item
the amplitude, $ \mybf{y_i - x_i \leq a} $,
\item
the Hamming weight, $ \mybf{\sum_{i=1}^n \mathbbm{1}_{\{y_i \neq x_i\}} \leq h} $,
and
\item
the total weight, $ \mybf{\sum_{i=1}^n (y_i - x_i) \leq t} $.
\end{inparaenum}
Optimal codes detecting these types of errors are described for certain sets of parameters
$ \mybf{a, h, t} $, both in the standard and in the cyclic ($ \mybf{\operatorname{mod}\, q} $)
version of the problem.
It is also demonstrated that these codes are optimal in the large alphabet limit for
every $ \mybf{a, h, t} $ and every block-length $ \mybf{n} $.
\end{abstract}%


\vspace{-1mm}
\section{Introduction}
\label{sec:intro}

\IEEEPARstart{P}{hysical} properties of certain optical and flash memory devices
that are used for information storage and transmission are such that the received
symbol (voltage, number of photons, etc.) can never be larger than the corresponding
transmitted symbol.%
\footnote{For convenience, we assume that the received symbol cannot be
\emph{smaller} than the transmitted one, which is of course equivalent.}
For this reason, the resulting communication models are usually referred to as
\emph{asymmetric} channels.
In the present paper we study the problem of error \emph{detection} in such channels.
In particular, we analyze the effect of the \emph{amplitude}, the \emph{Hamming weight},
and the \emph{total weight} of noise vectors on the size of optimal error-detecting codes.
This approach unifies and generalizes several known error models from the literature.
Our main results are proofs of optimality of a family of error-detecting codes for
some classes of channel parameters, and a proof that the same family is optimal in the
limit of large alphabets for all channel parameters and all block-lengths.

The paper is structured as follows.
In Section \ref{sec:model} we give a description of the type of asymmetric channels
we have in mind and the problem that will be analyzed in the sequel, as well as a
brief overview of the relevant literature.
Sections \ref{sec:infinite}, \ref{sec:finite}, \ref{sec:cyclic} contain our main
results concerning optimal error-detecting codes for asymmetric channels with infinite,
finite, and cyclic alphabets, respectively.
A brief conclusion and some pointers for further work are stated in Section \ref{sec:conclusion}. 

\vspace{-1mm}
\section{Model description and problem formulation}
\label{sec:model}

Let $ \mathbb{A} $ denote the channel alphabet, which we shall take to be
either $ \mathbb{Z} $---the set of all integers---or a subset of $ \mathbb{Z} $
of the form $ \{ 0, 1, \ldots, q - 1 \} $.
For any input vector $ {\bf x} = (x_1, \ldots, x_n) \in \mathbb{A}^n $, \pagebreak
the channel outputs a vector $ {\bf y} = (y_1, \ldots, y_n) \in \mathbb{A}^n $
satisfying the following conditions:
\begin{itemize}
\item[1)]
$ 0  \leq  y_i - x_i  \leq  a $,
\item[2)]
$ \sum_{i=1}^n \mathbbm{1}_{\{y_i \neq x_i\}} \leq h $,
\item[3)]
$ \sum_{i=1}^n (y_i - x_i) \leq t $.
\end{itemize}
Hence, we consider asymmetric channels with additional constraints imposed on:
\begin{inparaenum}
\item[1)]
the amplitude of the noise at each coordinate (so-called limited-magnitude errors),
\item[2)]
the number of symbols hit by noise, i.e., the Hamming weight of the error
vector $ {\bf y} - {\bf x} $, and
\item[3)]
the total weight of the error vector.
\end{inparaenum}
The error vectors satisfying the above constraints will be referred to as
$ (a, h, t) $-asymmetric errors.
Note that the situations where only some of the above three constraints are
imposed on the noise are special cases of our setting.
Namely, by taking $ a = t $ the constraint 1) is effectively excluded (i.e.,
becomes redundant), as it is when $ a = q - 1 $ in the finite alphabet case.
Similarly, one can exclude constraint 2) by setting $ h = n $, and
constraint 3) by setting $ t = a h $.
We shall refer to the $ (q-1, n, t) $-asymmetric errors (when only the constraint
3) is in effect) as the $ (\cdot, \cdot, t) $-asymmetric errors, and similarly
for the other cases.
Hence, the `$ \cdot $' indicates that the corresponding constraint is either
redundant, or is not being considered at all.

\textit{\textbf{Convention:}
To avoid discussing trivial cases, as well as to simplify the exposition, we
shall assume hereafter that $ q, n, a, h, t $ are positive integers satisfying
$ a \leq q - 1 $ (in the finite alphabet case), $ a \leq t \leq a h $, and $ h \leq n $.}
\myqed

\begin{figure}%[h]
\centering
\subfloat[$ (2, 1, \cdot) $-asymmetric errors.]
{
  \centering
  \includegraphics[width=0.38\columnwidth]{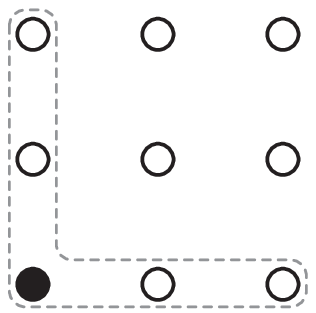}
  \label{fig:region_a2_h1}
}
\hspace{5ex}
\subfloat[$ (\cdot, \cdot, 2) $-asymmetric errors.]
{
  \centering
  \includegraphics[width=0.38\columnwidth]{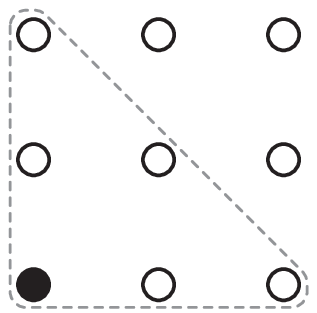}
  \label{fig:region_t2}
}\\
\subfloat[$ (2, \cdot, 3) $-asymmetric errors.]
{
  \centering
  \includegraphics[width=0.38\columnwidth]{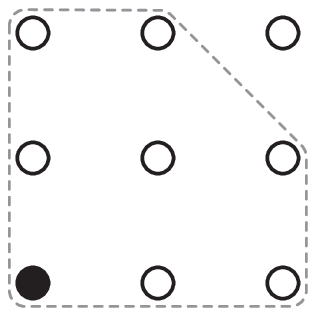}
  \label{fig:region_a2_t3}
}
\hspace{5.7ex}
\subfloat[$ (2, \cdot, \cdot) $-asymmetric errors.]
{
  \centering
  \includegraphics[width=0.38\columnwidth]{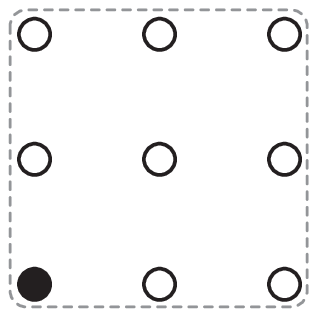}
  \label{fig:region_a2}
}
\caption{The region $ \out{0} $---the set of channel outputs that can be produced by the
input vector $ \bf 0 $ (depicted as black dot)---for various channel parameters and
block-length $ n = 2 $.}%
\label{fig:out0}
\end{figure}%

For $ {\bf x} \in \mathbb{A}^n $, denote by $ \out{x} $ the set of all channel
outputs $ {\bf y} \in \mathbb{A}^n $ that can be produced by the input $ {\bf x} $
and the noise vectors satisfying the constraints 1)--3) (see Figure \ref{fig:out0}).
The dependence of $ \out{x} $ on the parameters $ a, h, t $ is suppressed
for notational simplicity; this should not cause any confusion.

A code $ {\mathcal C} \subseteq \mathbb{A}^n $ is said to detect certain error
patterns if no codeword $ {\bf x} $ can produce another codeword $ {\bf y} \neq {\bf x} $
at the channel output, when any of those error patterns occur.
In symbols, if $ {\bf x} \in {\mathcal C} $ and $ {\bf y} \in \out{x} $,
$ {\bf y} \neq {\bf x} $, then $ {\bf y} \notin {\mathcal C} $.
This ensures that the received vector is either the transmitted codeword, or
not a codeword at all, meaning that the receiver can unambiguously determine
whether an error has happened during transmission.
Our object of study in this paper are codes detecting $ (a, h, t) $-asymmetric
errors;%
\footnote{Codes detecting asymmetric errors can also detect so-called unidirectional
errors---errors which are either positive or negative within a codeword, but the sign
is not known in advance; see, e.g., \cite{bose+lin}.}
they will be referred to as $ (a, h, t) $-asymmetric-error-detecting codes, or
$ (a, h, t) $-AED codes for short.

\vspace{-1mm}
\subsection*{Related work}

The work most closely related to ours, in which optimal $ (\cdot, \cdot, t) $-AED
codes over $ q $-ary alphabets were determined, is \cite{borden}.
In fact, we show that the same family of codes that was studied there remains
optimal in the case of $ (a, h, t) $-asymmetric errors in some instances.
We also extend and generalize the results of \cite{borden} to infinite alphabets.

The mentioned construction from \cite{borden} was also used in \cite{ahlswede1}
for the $ (a, \cdot, \cdot) $ case, and a very similar construction (also for the
$ (a, \cdot, \cdot) $ case, but in a quite different setting) appears in \cite{kovacevic+stojakovic+tan}.
Systematic $ (a, \cdot, \cdot) $-AED codes (as well as $ (a, h, \cdot) $-AED codes) were studied in \cite{elarief}.

Some other works on the error detection problem in asymmetric channels should
also be mentioned, e.g., \cite{albassam+bose, bose+elmougy+tallini, bose+lin,
bose+pradhan, bose+rao, naydenova+klove}.
All of these works are focused on special cases of the model introduced above---%
$ (\cdot, \cdot, t) $, $ (a, h, \cdot) $, binary alphabet, etc.
In \cite{tallini+bose}, a generalization of the $ (\cdot, \cdot, t) $ error model
is studied where both positive and negative errors are allowed.

Finally, for a study of the error \emph{correction} problem in asymmetric channels
we refer the reader to \cite{klove}; see also \cite{cassuto, klove+bose+elarief,
kloveetal, schwartz, yari+klove+bose} for some of the more recent works.

\section{Asymmetric channels with infinite alphabet}
\label{sec:infinite}

We first consider the case when the channel alphabet is the set of all integers
$ \mathbb{Z} $.
Though this is clearly not practically motivated, there are several reasons why
these results are relevant for the study of communication over asymmetric channels.
First, this will provide geometric intuition about the problem and allow us to
``visualize'' AED codes.
Second, the codes described below will be shown optimal for all channel parameters
in the infinite alphabet case, suggesting that the corresponding codes in the finite
alphabet case are nearly optimal, at least in some asymptotic regimes.
In fact, as we already mentioned, they will be proven optimal in some special
instances of the finite alphabet case as well.
Third, infinite alphabet can be seen as a limiting case and an approximation of
a finite alphabet; this is relevant in situations where the alphabet size $ q $ is
large compared to the ``error radius'' $ t $.

Before stating the results we need a few definitions.
To quantify what it means for a code in $ \mathbb{Z}^n $ to be optimal,
we define the density of $ {\mathcal C} \subseteq \mathbb{Z}^n $ as follows:
\begin{equation}
\label{eq:density}
  \mu({\mathcal C})  \defeq  \lim_{k \to \infty}
    \frac{ \left| {\mathcal C} \cap \{-k, \ldots, k\}^n \right| }{ (2k+1)^{n} } .
\end{equation}
This parameter represents the infinite-space analog of the cardinality of
codes in finite spaces.
In case the limit in \eqref{eq:density} does not exist, one can naturally define
the upper ($ \overline{\mu}({\mathcal C}) $) and the lower
($ \underline{\mu}({\mathcal C}) $) density by replacing $ \lim $ with
$ \limsup $ and $ \liminf $, respectively.
We say that $ {\mathcal C} $ is an optimal $ (a, h, t) $-AED code in $ \mathbb{Z}^n $
if no $ (a, h, t) $-AED code in $ \mathbb{Z}^n $ has upper density larger than
$ \overline{\mu}({\mathcal C}) $.
A code $ {\mathcal C} \subseteq \mathbb{Z}^n $ is said to be linear if it is a sublattice
of $ \mathbb{Z}^n $, i.e., if $ ({\mathcal C},+) $ is a subgroup of $ (\mathbb{Z}^n,+) $.
For a linear code we have
$ \mu({\mathcal C}) = \frac{1}{|\mathbb{Z}^n / {\mathcal C}|} $, where
$ \mathbb{Z}^n / {\mathcal C} $ is the quotient group of the lattice $ \mathcal C $.

For $ S \subset \mathbb{Z}^n $, we say that $ (S, {\mathcal C}) $ is a
\emph{packing} in $ \mathbb{Z}^n $ if the translates $ {\bf x} + S $ and $ {\bf y} + S $
are disjoint for any two distinct codewords $ {\bf x}, {\bf y} \in {\mathcal C} $
(here $ {\bf x} + S = \{{\bf x} + {\bf s} : {\bf s} \in S \} $).
If $ (S, {\mathcal C}) $ is a packing and $ {\bf 0} \in S $, then each of these translates
contains exactly one codeword, and so we must have $ \overline{\mu}({\mathcal C}) \leq \frac{1}{|S|} $.

The following claim gives an upper bound on the density of $ (a, h, t) $-AED
codes in $ \mathbb{Z}^n $.

\begin{theorem}
\label{thm:density}
  Let $ {\mathcal C} $ be an $ (a, h, t) $-AED code in $ \mathbb{Z}^n $.
Then $ \overline{\mu}({\mathcal C}) \leq \frac{1}{t+1} $.
\end{theorem}
\begin{IEEEproof}
  Let $ {\bf e}_i $ be the unit vector having a $ 1 $ at the $ i $'th coordinate
and $ 0 $'s elsewhere.
Observe the following vectors:
%\begin{equation}
%\label{eq:list}
 %\begin{aligned}
  %& {\bf 0}  \\
	%& {\bf e}_1  \\
	%& 2 \cdot {\bf e}_1  \\
	%& \;\; \vdots  \\
	%& a \cdot {\bf e}_1  \\
	%& a \cdot {\bf e}_1 + {\bf e}_2  \\
	%& \;\; \vdots  \\
	%& a \cdot {\bf e}_1 + a \cdot {\bf e}_2  \\
	%& \;\; \vdots  \\
	%& a \cdot {\bf e}_1 + a \cdot {\bf e}_2 + \cdots + a \cdot {\bf e}_{k-1} + \alpha_k \cdot {\bf e}_k
 %\end{aligned}
%\end{equation}
\begin{equation}
\label{eq:list}
 \begin{aligned}
  & {\bf 0}, \quad {\bf e}_1, \quad 2 \cdot {\bf e}_1, \quad \ldots, \quad a \cdot {\bf e}_1,  \\
	& a \cdot {\bf e}_1 + {\bf e}_2, \quad \ldots, \quad a \cdot {\bf e}_1 + a \cdot {\bf e}_2,  \\
	& \;\; \vdots  \\
	& a \cdot {\bf e}_1 + a \cdot {\bf e}_2 + \cdots + a \cdot {\bf e}_{k-1} + \alpha_k \cdot {\bf e}_k
 \end{aligned}
\end{equation}
where the list extends until one of the constraints on the noise is violated.
In other words, the last vector on the list is of the form
$ \sum_{i=1}^{h} \alpha_i \cdot {\bf e}_i $, where, for some $ k \in \{1, \ldots, h\} $,
$ \alpha_i = a $ for all $ i < k $, $ \alpha_i = 0 $ for all $ i > k $, and
$ \sum_{i=1}^k \alpha_i = t $.
Denote the set of all vectors on the resulting list by $ S $, and note that $ |S| = t + 1 $.
An important observation about this set is that, for any two vectors $ {\bf f}, {\bf g} \in S $,
where $ \bf f $ precedes $ \bf g $ on the list \eqref{eq:list}, the vector $ {\bf g} - {\bf f} $
satisfies all the noise constraints:
$ {\bf g} - {\bf f} = \sum_{i=1}^{h} \gamma_i \cdot {\bf e}_i $ with $ 0 \leq \gamma_i \leq a $
and $ \sum_{i=1}^{h} \gamma_i \leq t $.
This means that, for any two such vectors $ {\bf f}, {\bf g} $ we must have
$ {\bf g} - {\bf f} \in \out{0} $.\\
Now, let $ {\mathcal C} \subseteq \mathbb{Z}^n $ be an $ (a, h, t) $--AED code,
$ |{\mathcal C}| \geq 2 $.
We claim that $ (S, {\mathcal C}) $ is a packing in $ \mathbb{Z}^n $.
Suppose that this is not the case, i.e., that $ {\bf x} + {\bf f} = {\bf y} + {\bf g} $
for two distinct codewords $ {\bf x}, {\bf y} \in {\mathcal C} $ and two distinct
vectors $ {\bf f}, {\bf g} \in S $.
Without loss of generality, we can assume that $ \bf f $ precedes $ \bf g $ on the
list \eqref{eq:list}.
We then have $ {\bf g} - {\bf f} \in \out{0} $ and so
$ {\bf x} = {\bf y} + {\bf g} - {\bf f} \in \out{y} $, which means that the code
$ \mathcal C $ is not $ (a, h, t) $-AED, a contradiction.
Therefore, any $ (a, h, t) $-AED code defines a packing of the set $ S $ in
$ \mathbb{Z}^n $, and so its density cannot exceed $ \frac{1}{|S|} = \frac{1}{t+1} $.
\end{IEEEproof}

\begin{figure}%[h]
 \centering
  \includegraphics[width=0.96\columnwidth]{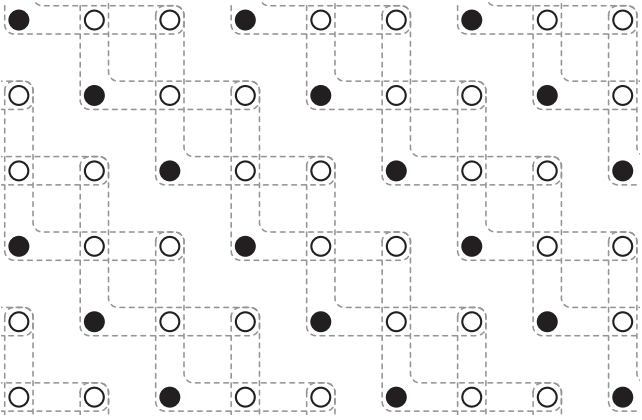}
\caption{The code $ {\mathcal C}(2;2) \subseteq \mathbb{Z}^2 $ and an illustration of
         its $ (2,1,\cdot) $-AED property. Codewords are depicted as black dots.}
\label{fig:code}
\end{figure}

We next give an explicit construction of \emph{linear} codes achieving the upper
bound just derived.
A particular such code is depicted in Figure \ref{fig:code}.

%\pagebreak
\begin{theorem}
\label{thm:optimalcodeZ}
  The code
\begin{equation}
\label{eq:code}
  {\mathcal C}(n;t)  \defeq  \left\{ {\bf x} \in \mathbb{Z}^n : \sum_{i=1}^n x_i \equiv 0  \mod (t	 + 1) \right\}
\end{equation}
is an optimal $ (a, h, t) $-AED code in $ \mathbb{Z}^n $.
\end{theorem}
\pagebreak
\begin{IEEEproof}
  Due to our assumptions about the noise, each of the allowed error vectors from
$ \out{0} $ can change the sum of the codeword symbols, $ \sum_{i=1}^n x_i $, by
at most $ t $.
Therefore, no codeword of $ {\mathcal C}(n;t) $, other than the one that was
transmitted, can be produced at the output of the channel, proving that this
code is indeed $ (a, h, t) $-AED.
To demonstrate its optimality, observe that the quotient group of the lattice
$ {\mathcal C}(n;t) $ is $ \mathbb{Z}^n / {\mathcal C}(n;t) \cong \mathbb{Z}_{t+1} $.
The density of $ {\mathcal C}(n;t) $ is therefore
$ \mu\left({\mathcal C}(n;t)\right) = \frac{1}{|\mathbb{Z}_{t+1}|} = \frac{1}{t+1} $,
which is by Theorem \ref{thm:density} the largest possible value.
\end{IEEEproof}

The code $ {\mathcal C}(n;t) $ can also be written in the form
$ {\mathcal C}(n;t) = \big\{ \boldsymbol{\xi} \cdot {\bf G}(n;t) : \boldsymbol{\xi} \in \mathbb{Z}^n \big\} $,
where $ {\bf G}(n;t) $ is an $ n \times n $ generator matrix:
\begin{equation}
 {\bf G}(n;t) =
    \begin{pmatrix}
       t+1      &  0       &  0       &  \cdots  &  0       \\
       -1       &  1       &  0       &  \cdots  &  0       \\
       -1       &  0       &  1       &  \cdots  &  0       \\
      \vdots    &  \vdots  &  \vdots  &  \ddots  &  \vdots  \\
       -1       &  0       &  0       &  \cdots  &  1
    \end{pmatrix}.
\end{equation}
%
%The generator matrix of the dual code $ {\mathcal C}^*(n;t) $ is
%\begin{equation}
 %H(n;t) = G(n;t)^{-\textsc{t}} =
    %\begin{pmatrix}
       %\frac{1}{t+1}    &  \frac{1}{t+1}   &  \frac{1}{t+1}   &  \cdots  &  \frac{1}{t+1}   \\
       %0      &  1       &  0       &  \cdots  &  0       \\
       %0      &  0       &  1       &  \cdots  &  0       \\
      %\vdots  &  \vdots  &  \vdots  &  \ddots  &  \vdots  \\
       %0      &  0       &  0       &  \cdots  &  1
    %\end{pmatrix} .
%\end{equation}

\begin{remark}
  Note that the codes $ {\mathcal C}(n;t) $---optimal $ (a, h, t) $-AED codes in
$ \mathbb{Z}^n $---do not depend on the parameters $ a, h $, a somewhat counter-%
intuitive fact.
Hence, detecting $ (\cdot, \cdot, t) $-asymmetric errors incurs no loss in code
efficiency compared to the case of detecting more restrictive $ (a, h, t) $-%
asymmetric errors.
The corresponding statement for error-\emph{correcting} codes is in general false.
Note that the density of $ {\mathcal C}(n;t) $ is independent of the block-length
$ n $ as well.%
\footnote{In the finite alphabet case, it is known that the number of redundant
symbols of an optimal $ (\cdot, \cdot, t) $-AED code depends only on $ t $ and not on
the block-length $ n $, see for example \cite{borden, bose+elmougy+tallini, bose+lin}.}
\myqed
\end{remark}

\section{Asymmetric channels with finite alphabet}
\label{sec:finite}

The main idea in constructing codes over finite alphabets is simple: take an
$ (a, h, t) $-AED code in $ \mathbb{Z}^n $ and restrict it to the hypercube
$ \{0, 1, \ldots, q-1\}^n $.
This will clearly yield an $ (a, h, t) $-AED code with alphabet
$ [q] \defeq \{0, 1, \ldots, q-1\} $.
We say that $ {\mathcal C} $ is an optimal $ (a, h, t) $-AED code in $ [q]^n $
if it 	has the largest cardinality among all $ (a, h, t) $-AED codes in $ [q]^n $.

Since the code $ {\mathcal C}(n;t) $ from \eqref{eq:code} is optimal in $ \mathbb{Z}^n $,
it is natural to take it, or any of its translations, as the basis for construction.
In other words, we consider codes in $ [q]^n $ of the form
$ \left( {\bf z} + {\mathcal C}(n;t) \right) \cap [q]^n $,
for an arbitrary vector $ {\bf z} \in \mathbb{Z}^n $.
The resulting family of codes can be written as
\begin{equation}
\label{eq:codeq}
  {\mathcal C}_q^{(j)}(n;t)  \defeq  \left\{ {\bf x} \in [q]^n : \sum_{i=1}^n x_i \equiv j \mod (t + 1) \right\} ,
\end{equation}
where $ j \in \{0, 1, \ldots, t\} $.
The cardinality of $ {\mathcal C}_q^{(j)}(n;t) $ is maximized (with respect to $ j $)
for $ j = j^* \defeq \lfloor \frac{(q-1)n}{2} \rfloor \mymod (t+1) $, see \cite{borden}.

The codes $ {\mathcal C}_q^{(j^*)}(n;t) $ are known \cite{borden} to be optimal
$ (\cdot,\cdot,t) $-AED codes for every $ q, n, t $.
We prove below their optimality in some other cases as well.
The proof method from \cite{borden}, however, does not seem to be applicable to
these cases due to a different shape of the regions $ \out{x} $.

\begin{theorem}
\label{thm:densityq}
  Let $ {\mathcal C} $ be a $ (\cdot,h,t) $-AED code in $ [q]^n $.
Then $ |{\mathcal C}|  \leq  q^{n-1} \lceil \frac{q}{t+1} \rceil $.
\end{theorem}
\begin{IEEEproof}
  Partition the space $ [q]^n $ into $ q^{n-1} $ ``lines'', each containing $ q $
points whose coordinates $ 2,\ldots,n $ are fixed and the first coordinate
varies through $ [q] $.
If $ {\bf x} = (x_1, x_2, \ldots, x_n) $ is a codeword of a $ (\cdot,h,t) $-AED code,
then $ (x_1 + k, x_2, \ldots, x_n) $ cannot be a codeword for any $ 1 \leq k \leq t $,
because $ (x_1 + k, x_2, \ldots, x_n) \in \out{x} $.
This implies that each of the mentioned lines contains at most $ \lceil \frac{q}{t+1} \rceil $
codewords, and therefore $ |{\mathcal C}|  \leq  q^{n-1} \lceil \frac{q}{t+1} \rceil $.
\end{IEEEproof}

\begin{theorem}
\label{thm:optimalcodeq}
  Suppose that $ t + 1 $ divides $ q $.
Then the codes $ {\mathcal C}_q^{(j)}(n;t) $ are optimal $ (\cdot,h,t) $-AED codes in $ [q]^n $.
Their cardinality is $ \big| {\mathcal C}_q^{(j)}(n;t) \big| =  \frac{q^n}{t+1} $,\;
$ \forall	 j \in \{0, 1, \ldots, t\} $.
\end{theorem}
\begin{IEEEproof}
  For every choice of the values $ x_1, \ldots, x_{n-1} \in [q] $ there are
$ \frac{q}{t+1} $ possible values of $ x_n \in [q] $  satisfying the
congruence $ \sum_{i=1}^n x_i \equiv j \mymod (t+1) $, for any fixed $ j $.
Therefore, $ |{\mathcal C}_q^{(j)}(n;t)| = q^{n-1} \cdot \frac{q}{t+1} $ for
all $ j \in \{0, 1, \ldots, t\} $, which is by Theorem \ref{thm:densityq}
the largest possible value.
\end{IEEEproof}

The following statement refers to codes detecting a single ($ h = 1 $)
asymmetric error, with no bound on the amplitude of the error, other than the
implicit one $ a \leq q - 1 $.
It is a special case of Theorem \ref{thm:optimalcodeq} but we state it
separately nonetheless because the corresponding problem for error\linebreak
correction has been studied in some detail in the literature \cite{klove+bose+elarief, kloveetal, zhang+ge}.

\begin{corollary}
  The codes $ {\mathcal C}_q^{(j)}(n;q-1) $ are optimal $ (\cdot,1,\cdot) $-AED
codes in $ [q]^n $.
Their cardinality is $ \big| {\mathcal C}_q^{(j)}(n; q-1) \big| =  q^{n-1} $,
$ \forall	 j \in \{0, 1, \ldots, q-1\} $.
\end{corollary}
\begin{IEEEproof}
  Take $ t = q - 1 $ in Theorem \ref{thm:optimalcodeq} and notice that
$ (\cdot, 1, q - 1) $-asymmetric errors are in fact $ (\cdot, 1, \cdot) $-%
asymmetric errors.
\end{IEEEproof}

It should be noted that the codes $ {\mathcal C}_q^{(j)}(n;t) $ are not
optimal for general $ (a, h, t) $-asymmetric errors---counterexamples can
be constructed for small values of these parameters (see \cite{ahlswede1}
for a counterexample for the $ (a, \cdot, \cdot) $ case).

\section{Asymmetric channels with cyclic alphabet}
\label{sec:cyclic}

In this section we discuss briefly the cyclic version of the asymmetric channel
(as in, e.g., \cite{klove+bose+elarief, yari+klove+bose}).
Our motivating example is the so-called noisy typewriter channel wherein each
transmitted symbol $ x_i $ is received as either $ x_i $, or $ x_i + 1 \mymod q $.

Let $ +_q $ denote addition modulo $ q $.
The cyclic asymmetric channel we have in mind is defined as follows:
for any input vector $ {\bf x} \in [q]^n $ the channel outputs $ {\bf y} = {\bf x} +_q {\bf f} \in [q]^n $,
where $ {\bf f} \in [q]^n $ is an arbitrary noise vector satisfying the constraints
1)--3) described in Section \ref{sec:model}.
Hence, the model is the same as before, the only difference being that the sum
of the input vector and the noise vector is now taken $ \operatorname{mod}\, q $;
in other words, we now allow the errors to ``wrap around''.
To distinguish between cyclic and non-cyclic cases, we shall refer to the errors
just described as $ (a, h, t)^\circ $-asymmetric errors, and similarly for the
corresponding codes.

The code space in this setting can be represented as the torus $ \mathbb{Z}_q^n $
in which there are no ``boundary effects'' that are present in the non-cyclic case.
This enables one to derive a simple upper bound on the cardinality of optimal
codes by using a method identical to the one used for the infinite alphabet
case.

\begin{theorem}
\label{thm:densitymodq}
  Let $ {\mathcal C} $ be an $ (a, h, t)^\circ $-AED code in $ [q]^n $.
Then $ |{\mathcal C}| \leq \frac{q^n}{t+1} $.
\end{theorem}
\begin{IEEEproof}
  Analogous to the proof of Theorem \ref{thm:density}.
\end{IEEEproof}

We next identify a class of parameters for which the above bound is tight.
(As we shall point out in Section \ref{sec:conclusion}, it cannot be tight in
general.)

\begin{theorem}
\label{thm:optimalcodemodq}
  Suppose that $ t + 1 $ divides $ q $.
Then the codes $ {\mathcal C}_q^{(j)}(n;t) $ are optimal $ (a, h, t)^\circ $-AED codes in $ [q]^n $.
\end{theorem}
\begin{IEEEproof}
  We have shown in Theorem \ref{thm:optimalcodeq} that
$ \big| {\mathcal C}_q^{(j)}(n;t) \big| = \frac{q^n}{t+1} $ when $ t + 1 $ divides
$ q $, which is by Theorem \ref{thm:densitymodq} the maximum possible cardinality
of an $ (a, h, t)^\circ $-AED code in $ [q]^n $.
It is left to prove that $ {\mathcal C}_q^{(j)}(n;t) $ are indeed $ (a, h, t)^\circ $-AED.
We prove this fact below for $ j = 0 $; the statement for an arbitrary $ j $ is an
easy consequence.
The key observation is that, when $ t + 1 $ divides $ q $,
$ \big\{ {\bf x} \in \mathbb{Z}^n :
\exists {\bf x}' \in {\mathcal C}_q^{(0)}(n;t) \; \text{s.t.} \; {\bf x} \equiv {\bf x}' \mymod q \big\} =
{\mathcal C}(n;t) $, which follows from the definition of the codes $ {\mathcal C}(n;t) $
and $ {\mathcal C}_q^{(0)}(n;t) $.
In other words, the code $ {\mathcal C}(n;t) $ is a periodic extension to
$ \mathbb{Z}^n $ of the code $ {\mathcal C}_q^{(0)}(n;t) $.
With this interpretation in mind it is easy to see that the statement that $ {\mathcal C}_q^{(0)}(n;t) $
is $ (a, h, t)^\circ $-AED is equivalent to the statement that $ {\mathcal C}(n;t) $
is $ (a, h, t) $-AED, which we already know is true.
\end{IEEEproof}

\begin{corollary}
  The codes $ {\mathcal C}_q^{(j)}(n;q-1) $ are optimal $ (\cdot,1,\cdot)^\circ $-AED
codes in $ [q]^n $.
\hfill \IEEEQED
\end{corollary}

\section{Concluding remarks and further work}
\label{sec:conclusion}

The problem we have addressed in this letter is that of finding optimal error-%
detecting codes for asymmetric channels with various constraints on the noise.
The solution has been obtained in the infinite alphabet case for all parameters,
but for finite alphabets the general question is still open.
The task of settling it for every $ q, n, a, h, t $ may turn out to be too
difficult and it is instructive to focus on \emph{asymptotic} optimality instead.

For example, observe the regime where $ q, a, h, t $ are fixed and $ n \to \infty $.
Note that $ \big\{ {\mathcal C}_q^{(j)}(n;t) : j \in \{0, 1, \ldots, t\} \big\} $
is a partition of $ [q]^n $, meaning that the codes in this set are mutually disjoint
and their union is all of $ [q]^n $.
It is not difficult to argue that the members of this partition are of ``approximately
the same cardinality'' for large $ n $, and consequently
$ \big| {\mathcal C}_q^{(j)}(n;t) \big| \sim \frac{q^n}{t+1} $ for any
$ j \in \{0, 1, \ldots, t\} $
(here $ a_n \sim b_n $ is a shorthand for $ \lim_{n \to \infty} \frac{a_n}{b_n} = 1 $).
Whether this family of codes is asymptotically optimal is an interesting
question which we summarize below.
In light of Theorem \ref{thm:optimalcodeZ} one may conjecture that the answer
is positive.

{\it Problem:} Fix $ q, a, h, t $, and let $ D_{q}(n;a,h,t) $ denote the size
of an optimal $ (a, h, t) $-AED code in $ [q]^n $.
Is it true that $ D_{q}(n;a,h,t)  \sim  \frac{q^n}{t+1} $ as $ n \to \infty $?
\hfill \myqed

%\pagebreak
As for the cyclic case, we note that the cardinality of optimal codes cannot
scale as $ \frac{q^n}{t+1} $ in general.
For example, a code is $ (\cdot, h, \cdot)^\circ $-AED if and only if its
minimum Hamming distance is $ > h $, and it is known that such codes cannot
have size $ \sim c q^n $ for $ h \geq 2 $ (this follows from the sphere packing
bound in the $ q $-ary Hamming space).

\vfill
\section*{Acknowledgment}

The author would like to thank Vincent Y. F. Tan for reading a preliminary version
of this work and for several helpful discussions on the subject matter.
%, as well as the Associate Editor and the referees for their constructive comments.

\vfill
%\IEEEtriggeratref{9}

\end{document}